\documentclass{article}
\usepackage[utf8]{inputenc}
\usepackage{fancyhea}
\usepackage{bm}       
\usepackage{graphicx}
\usepackage{hyperref}      
\usepackage{amsmath}
\usepackage{cite}  
\usepackage{xcolor}

\newcommand{\nc}{\newcommand}  

\def\beq{\begin{equation}}
\def\eeq#1{\label{#1}\end{equation}}
\def\eeqn{\end{equation}}

\newenvironment{Eqnarray}%
   {\arraycolsep 0.14em\begin{eqnarray}}{\end{eqnarray}}
\def\beqa{\begin{Eqnarray}}
\def\eeqa#1{\label{#1}\end{Eqnarray}}
\def\eeqan{\end{Eqnarray}}

\nc{\ra}{\rightarrow}  
\nc{\slsh}{\slash\hspace*{-0.22cm}}
\def\Re{{\cal R \mskip-4mu \lower.1ex \hbox{\it e}\,}}
\def\Im{{\cal I \mskip-5mu \lower.1ex \hbox{\it m}\,}}

\nc{\vev}[1]{ \left\langle {#1} \right\rangle }
\nc{\bra}[1]{ \langle {#1} | }
\nc{\ket}[1]{ | {#1} \rangle }
\nc{\fb}{\,{\rm fb}^{-1}}
\nc{\ev}{{\rm eV}}
\nc{\kev}{{\rm keV}}
\nc{\Mev}{{\rm MeV}}
\nc{\gev}{{\rm GeV}}
\nc{\tev}{{\rm TeV}}
\nc{\mev}{{\rm MeV}}

\def\del{\partial}
\def\Dslash{\not{\hbox{\kern-4pt $D$}}}
\def\dslash{\not{\hbox{\kern-2pt $\del$}}}
\def\pslash{\not{\hbox{\kern-2pt $p$}}}
\def\ETmiss{ \not{\hbox{\kern-4pt $E$}}_T }

\def\msb{{\bar{\ssstyle M \kern -1pt S}}}

\usepackage{hyperref}
\usepackage{latexsym}
\usepackage{xspace}
\usepackage{relsize}
\usepackage{subfigure}
\usepackage[section]{placeins}

\usepackage[figuresright]{rotating}

\usepackage{amssymb}

\begin{document}
\pagenumbering{none}
\def\bibname{References}

\raggedbottom

\pagenumbering{arabic}

\thispagestyle{empty}

\parindent=0pt
\parskip=8pt





\centerline{\large A Design for an Electromagnetic Filter for Precision Energy Measurements}
\centerline{\large at the Tritium Endpoint}
 \noindent
M.G.~Betti$^{11,12}$,
M.~Biasotti$^{5}$,
A.~Bosc\'a$^{17}$,
F.~Calle$^{17}$,
J.~Carabe-Lopez$^{15}$,
G.~Cavoto$^{11,12}$,
C.~Chang$^{25,26}$,
W.~Chung$^{29}$,
A.G.~Cocco$^{7}$,
A.P.~Colijn$^{14}$,
J.~Conrad$^{20}$,
N.~D'Ambrosio$^{2}$,
P.F.~de~Salas$^{18,20}$,
M.~Faverzani$^{6}$,
A.~Ferella$^{20}$, 
E.~Ferri$^{6}$,
P.~Garcia-Abia$^{15}$,
G.~Garcia~Gomez-Tejedor$^{16}$,
S.~Gariazzo$^{18}$,
F.~Gatti$^{5}$,
C.~Gentile$^{28}$,
A.~Giachero$^{6}$,
J.E.~Gudmundsson$^{20}$,
Y.~Hochberg$^{1}$, 
Y.~Kahn$^{26,27}$,
M.~Lisanti$^{29}$,
C.~Mancini-Terracciano$^{11,12}$,
G.~Mangano$^{7}$,
L.E.~Marcucci$^{9,10}$,
C.~Mariani$^{12}$,
J.~Mart\'inez$^{17}$,
M.~Messina$^{22}$,
A.~Molinero-Vela$^{15}$,
E.~Monticone$^{13}$,
A.~Nucciotti$^{6}$,
F.~Pandolfi$^{11}$,
S.~Pastor$^{18}$,
J.~Pedr\'os$^{17}$,
C.~P\'erez~de~los~Heros$^{21}$,
O.~Pisanti$^{7,8}$,
A.~Polosa$^{11,12}$,
A.~Puiu$^{6}$,
Y.~Raitses$^{28}$,
M.~Rajteri$^{13}$,
N.~Rossi$^{11}$,
R.~Santorelli$^{15}$,
K.~Schaeffner$^{3}$,
C.F.~Strid$^{19,20}$,
C.G.~Tully$^{29}$,
F.~Zhao$^{29}$,
K.M.~Zurek$^{23,24}$

$^{1}$Racah Institute of Physics, Hebrew University of Jerusalem, Jerusalem, Israel\\
$^{2}$INFN Laboratori Nazionali del Gran Sasso, L'Aquila, Italy\\
$^{3}$Gran Sasso Science Institute (GSSI), L'Aquila, Italy\\
$^{4}$INFN Laboratori Nazionali di Frascati, Frascati, Italy\\
$^{5}$Universit\`a degli Studi di Genova e INFN Sezione di Genova, Genova, Italy \\
$^{6}$Universit\`a degli Studi di Milano-Bicocca e INFN Sezione di Milano-Bicocca, Milano, Italy\\
$^{7}$INFN Sezione di Napoli, Napoli, Italy \\
$^{8}$Universit\`a degli Studi di Napoli Federico II, Napoli, Italy\\
$^{9}$Universit\`a degli Studi di Pisa, Pisa, Italy \\
$^{10}$INFN Sezione di Pisa, Pisa, Italy\\
$^{11}$INFN Sezione di Roma, Roma, Italy \\
$^{12}$Sapienza Universit\`a  di Roma, Roma, Italy\\
$^{13}$Istituto Nazionale di Ricerca Metrologica (INRiM), Torino, Italy\\
$^{14}$Nationaal instituut voor subatomaire fysica (NIKHEF), Amsterdam, Netherlands\\
$^{15}$Centro de Investigaciones Energ\'eticas, Medioambientales y Tecnol\'ogicas (CIEMAT),  Madrid, Spain\\
$^{16}$Consejo Superior de Investigaciones Cientificas (CSIC), Madrid, Spain\\
$^{17}$Universidad Polit\'ecnica de Madrid, Madrid, Spain\\
$^{18}$Instituto de F\'{\i}sica Corpuscular  (CSIC-Universitat de Val\`{e}ncia), Valencia, Spain\\
$^{19}$Division of Physics, Lule{\aa} University of Technology, Lule{\aa}, Sweden\\
$^{20}$Stockholm University, Stockholm, Sweden\\
$^{21}$Uppsala University, Uppsala, Sweden\\
$^{22}$New York University Abu Dhabi, Abu Dhabi, UAE \\
$^{23}$Lawrence Berkeley National Laboratory, University of California, Berkeley, CA, USA \\
$^{24}$Department of Physics, University of California, Berkeley, CA, USA \\
$^{25}$Argonne National Laboratory, Chicago, IL, USA \\
$^{26}$Kavli Institute for Cosmological Physics, University of Chicago, Chicago, IL, USA \\
$^{27}$University of Illinois Urbana-Champaign, Urbana, IL, USA\\
$^{28}$Princeton Plasma Physics Laboratory, Princeton, NJ, USA \\
$^{29}$Department of Physics, Princeton University, Princeton, NJ, USA \\


\centerline{\bf Abstract}

We present a detailed description of the electromagnetic filter for the PTOLEMY project to directly detect the Cosmic Neutrino Background (CNB). 
Starting with an initial estimate for the orbital magnetic moment, 
the higher-order drift process of $\bm{E} \times \bm{B}$ is configured to balance the gradient-$B$ drift motion of the electron in such a way as to guide the trajectory into the standing voltage potential along the mid-plane of the filter.
As a function of drift distance along the length of the filter, the filter zooms in with exponentially increasing precision on the transverse velocity component of the electron kinetic energy.  This yields a linear dimension for the total filter length that is exceptionally compact compared to previous techniques for electromagnetic filtering.  The parallel velocity component of the electron kinetic energy oscillates in an electrostatic harmonic trap as the electron drifts along the length of the filter.  An analysis of the phase-space volume conservation validates the expected behavior of the filter from the adiabatic invariance of the orbital magnetic moment and energy conservation following Liouville's theorem for Hamiltonian systems.



\clearpage

\section*{Overview}
\addcontentsline{toc}{section}{Overview}

The concept of neutrino capture on $\beta$-decay nuclei as a detection method for the Cosmic Neutrino Background (CNB) was laid out in the original paper by Steven Weinberg~\cite{Weinberg:1962zza} in 1962 and further refined by the work of Cocco, Mangano and Messina~\cite{Cocco:2007za} in 2007 in view of the finite neutrino mass discovered by oscillation experiments. An experimental realization of the concept for CNB detection was proposed based on PTOLEMY~\cite{Betts:2013uya} in 2013 and a subsequent R\&D program described in~\cite{Baracchini2018}.  We describe in this paper the novel concept and principles of operation of the electromagnetic filter proposed for PTOLEMY for precision energy measurements at the tritium endpoint.

The filter concept builds on over half a century of electromagnetic filter techniques developed for the indirect estimation of the (anti-)neutrino\footnote{The Majorana or Dirac nature of the neutrino sector is, as of yet, unknown.} masses through the kinematic distortion on the phase space of electrons emitted near the tritium endpoint~\cite{hamilton1953upper,bergkvist1972high,lubimov1980estimate,Beamson1980Collimating,Lobashev1985Method,Picard1992Solenoid}.  
The original filters based on purely electrostatic filtering were enhanced through techniques of magnetic adiabatic collimation and have been developed into precision spectrometers~\cite{kraus2005final,aseev2011upper,wolf2010katrin}.  Collimation is a process where the transverse kinetic energy of an electron with respect to a magnetic field is transformed into a longitudinal kinetic energy, thereby aligning the total velocity of the electron along the magnetic field line.  This transformation is governed by the conditions of adiabatic invariance through the constancy of the orbital magnetic moment, $\mu$, as the electron motion follows magnetic field lines that expand from an initially high magnetic field region into low field.  This process trades the dense momentum phase space transverse to the magnetic field of the initial electron in gyromotion with a large transverse position phase space, as required by static Hamiltonian systems by Liouville's theorem.

The technique proposed here is a departure from previous approaches (referenced above); in that, the reduction of the electron velocity perpendicular to the magnetic field originates from the cyclotron-phase-averaged motion that is transverse to the magnetic field through higher-order drift processes.  Instead of collimating the transverse kinetic energy of the electron into a longitudinal velocity, the non-electric drift in the magnetic field gradient does work on the electron to reduce the transverse kinetic energy in favor of climbing a standing voltage potential.
Starting with an initial estimate for $\mu$, the higher-order drift process of $\bm{E} \times \bm{B}$ is configured to balance the gradient-$B$ drift motion of the electron in such a way as to guide the trajectory into the standing voltage potential along the mid-plane of the filter.
We show that this process, as expected from adiabatic invariance and energy conservation, conserves phase-space volume.

\section*{The Transverse Drift Filter Concept}
\addcontentsline{toc}{section}{The Transverse Drift Filter Concept}

An electron moving perpendicular to a magnetic field will, in general, undergo cyclotron motion, or gyromotion, from the Lorentz force.  We can describe the 
central axis of the trajectory of an electron in gyromotion with respect to a magnetic field by the guiding center system (GCS) variables formed by setting the cyclotron phase average of all forces acting on the particle to zero.  
The GCS is a non-inertial reference frame whose transverse plane is oriented orthogonal to the magnetic field direction\footnote{In what follows, we follow the terminology, conventions, and derivations given in this reference~\cite{roederer2014particle}.}.
The direction of the GCS trajectory, however, needs not to correspond to the direction of the magnetic field line and will, in general, deviate from the magnetic field direction in the presence of four fundamental drift terms:
\begin{equation} \label{eq:drifts}
\bm{V}_D = \bm{V}_\perp = \left( q\bm{E} + \bm{F} - \mu \bm{\nabla} B - m \frac{d\bm{V}}{dt} \right) \times \frac{\bm{B}}{qB^2} 
\end{equation}
where $q$ and $m$ are the electron charge and mass, respectively, and $\bm{V} = \bm{V}_\perp + \bm{V}_\parallel$ is the total phase-averaged velocity of the GCS trajectory with perpendicular and parallel components with respect to the magnetic field line.  The perpendicular component is often referred to as the drift velocity, $\bm{V}_D$.  In equation~(\ref{eq:drifts}), the four drift terms, from left to right, are given by (1) the $\bm{E} \times \bm{B}$ drift; (2) the external force drift (such as gravity); (3) the gradient-$B$ drift; and (4) the inertial force drift.

The GCS description is valid in the limit that the $E$ and $B$ fields vary slowly spatially relative to the cyclotron radius, $\rho_c$, and slowly in time, through the motion of the particle, compared to the cyclotron
period, $\tau_c$, namely:
\begin{align} \label{eq:adinv1}
\rho_c & \ll \left| \frac{B}{\nabla{B}} \right|, \ \left| \frac{E}{\nabla{E}} \right| \ ; \ \mathrm{and} \\
\tau_c & \ll \left| \frac{B}{dB/dt}\right|, \ \left|\frac{E}{dE/dt}\right| \ ; \label{eq:adinv2}
\end{align}
where the total variation per unit time seen by the particle comes from the variation in time at a fixed point in space and the variation due to the displacement while the field is fixed in time: $d/dt = \partial/\partial t + \bm{V \cdot \nabla}$.
These conditions, if satisfied, allow the motion of the electron to be accurately described by adiabatic invariants, and, in particular, the first adiabatic invariant.  The derivation of the first adiabatic invariant
is found in these references~\cite{Alfven1940,cary2009hamiltonian}
and follows from the action-angle variable description of the Hamiltonian in 
terms of the gyroaction $J \equiv (mc/q)\mu$ canonically conjugate to the cyclotron phase angle, where $\bm{\mu}$, with magnitude $\mu$, is the
orbital magnetic moment of the electron with respect to a magnetic field $\bm{B}$.
Starting with a non-relativistic treatment, $\mu$ in the GCS frame is given by
\begin{equation}
\mu = \frac{m v_\perp^{*2}}{2 B}
\end{equation}
where $\bm{v}_\perp^*$ is the instantaneous velocity of the electron
perpendicular to the magnetic field line in the GCS frame (starred quantities) and are related to the inertial frame 
instantaneous velocity $\bm{v} = \bm{v}_\perp + \bm{v}_\parallel$ by
$\bm{v}_\perp^* = \bm{v}_\perp - \bm{V}_D$ and
$\bm{v}_\parallel^* = \bm{v}_\parallel - \bm{V}_\parallel \approx 0$.
The angle, $\alpha$, between $\bm{v}$ and $\bm{B}$, also equal to
\begin{equation}
\alpha = \arccos \frac{v_\parallel}{v} \ ,
\end{equation}
is the {\it pitch angle} of the electron.

In the presence of a non-uniform magnetic field, the Hamiltonian term 
$U = - \bm{ \mu \cdot B}$ gives rise to a total net force given by
\begin{equation}
\bm{f} = - \bm{\nabla} U = - \mu \bm{\nabla} B \ .
\end{equation}
The parallel component, $f_\parallel$,
is the well-known {\it mirror force} responsible for magnetic adiabatic 
collimation and the magnetic bottle effect for trapping charged particles in
non-uniform magnetic fields.
The perpendicular component, $f_\perp$, is the source of the gradient-$B$ drift.
This drift is particularly interesting for a filter since only non-electric
drifts can lead to a change in total kinetic energy.  Drifts due to electric fields are always perpendicular to $\bm{E}$ by construction and therefore cannot do any work -- electrons under $\bm{E} \times \bm{B}$ drift follow surfaces of constant voltage.

More precisely, when accompanied by $\bm{E} \times \bm{B}$ drift, the gradient-$B$ drift
can do work on the electron and reduce the internal kinetic energy of gyromotion 
for a corresponding increase in voltage potential.  This is described by, inserting terms from equation~(\ref{eq:drifts}), 
\begin{equation} \label{eq:energycon}
\frac{d T_\perp}{dt} = - q \bm{E \cdot V}_D = - q \bm{E \cdot}
\left( q\bm{E} - \mu \bm{\nabla} B \right) \times \frac{\bm{B}}{qB^2}  
= \frac{\mu}{B^2} \bm{E \cdot} ( \bm{\nabla} B \times \bm{B} ) 
\end{equation}
where $T_\perp$ is the internal kinetic energy of gyromotion in the GCS frame.
The implementation of this basic principle into a filter for PTOLEMY is described below.

\section*{Implementation of PTOLEMY Filter}
\addcontentsline{toc}{section}{Implementation of the PTOLEMY Filter}

\begin{figure*}[h!]
\begin{center}
      \includegraphics[width=0.95\textwidth]{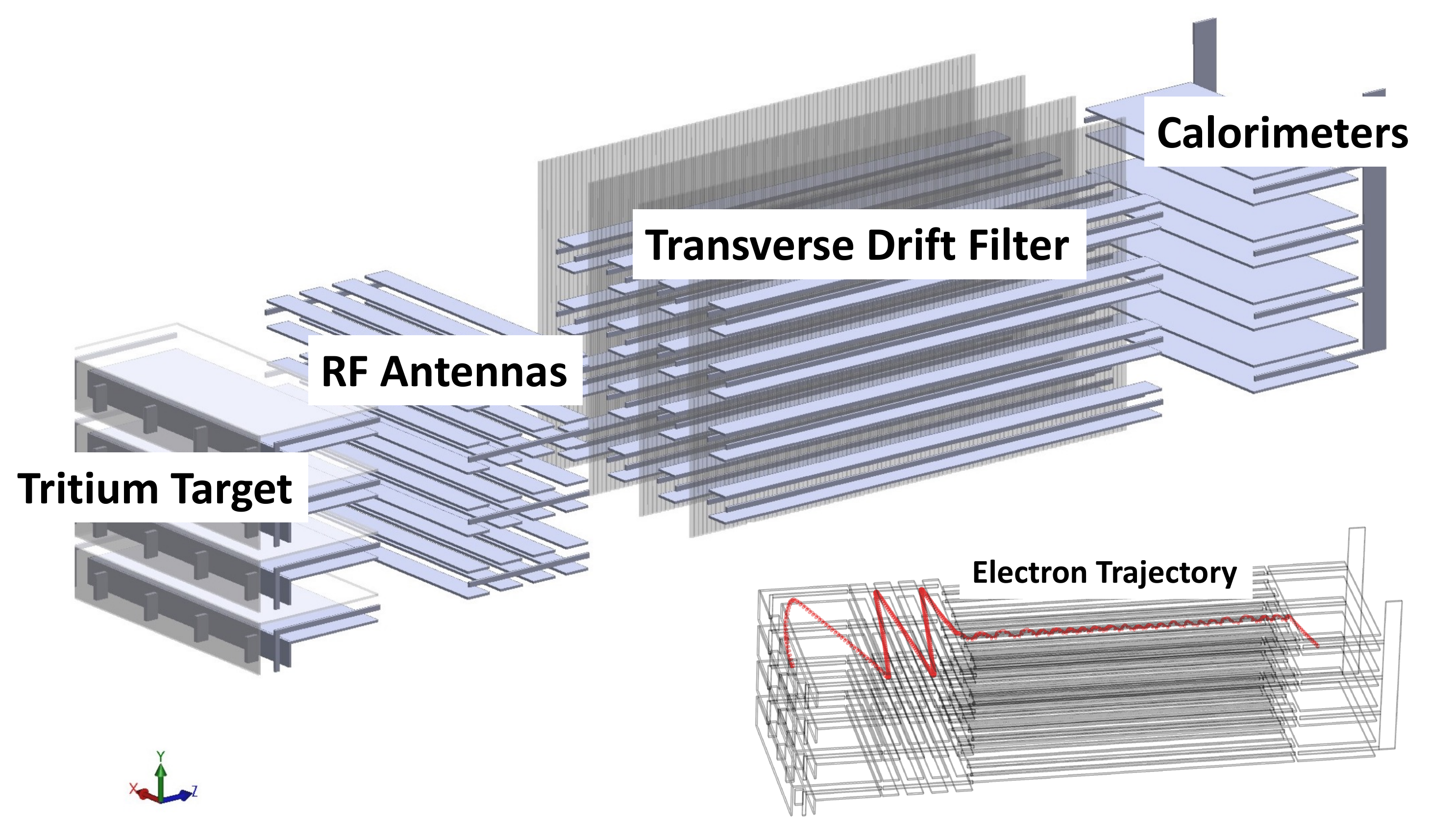}
   \caption{Diagram of the PTOLEMY layout.  Electrons originate in the tritium target, drift through the RF antennas and enter the transverse drift filter.  An example electron trajectory is shown ({\it red}) in the lower right corner, computed with COMSOL software~\cite{multiphysics2015v}.
The transverse drift filter consists of a set of configurable electrodes to setup the electric fields for $\bm{E} \times \bm{B}$ drift and a set of planar geometry current coils (vertical planes) to provide a magnetic field gradient in the drift region.  The filter element is configured to transport electrons near the tritium endpoint down the central axis of the filter, to within the accuracy of the estimate of $\mu$ in advance of entering the filter.  At the exit of the filter, the electron will be guided into a calorimeter.
   }
\label{fig:PTOLEMYfilter}
\end{center}
\vspace*{-15pt}
\end{figure*}

The implementation of this technique in PTOLEMY to study the tritium endpoint is described in the following sections.  We assume an estimate for the transverse kinetic energy from the measurement of the cyclotron radiation emission of single electrons in gyromotion developed by Project~8~\cite{project8}.  This estimate is derived from the radio-frequency (RF) signal emitted by the electron during the slow transport of the electron in a region of constant magnetic field in advance of the filter, assuming low occupancy in the filter for electrons near the tritium endpoint.  Depending on the measured value of the transverse kinetic energy, the voltage levels in the filter are adjusted in advance of the electron entering the filter so that for a given initial transverse kinetic energy and a starting voltage level, the electron drifts along a fixed trajectory through the filter to within the accuracy of the initial RF measurement and precision of the voltage steps.  We assume for the sake of this paper that the duration of the electron transport through the RF antenna system, which can be adjusted by lowering the $\bm{E} \times \bm{B}$ drift velocity in that region, is sufficiently long, on order a millisecond, to provide ample time for the voltages in the filter to settle before the electron enters the filter.  At the exit of the filter, the electron is guided into a calorimeter to complete the precision energy measurement of the total initial kinetic energy of the electron emitted from the tritium nuclei.  The level of precision and accuracy required from the RF estimate is set by the dynamic range of the calorimeter and is assumed here to be a few eV.
A diagram of the PTOLEMY layout is shown in Figure~\ref{fig:PTOLEMYfilter}.

The precision energy measurement of electrons from the tritium endpoint is primarily a combination of two measurements, the reference electrical potential energy difference from the tritium target to the calorimeter and the calorimetric measurement of the total kinetic energy of the electron at the exit of the filter.  Residual energy corrections are applied to account for radio-frequency energy losses during transport.  The target resolution of the combined measurement is 0.05\,eV, corresponding to the largest neutrino mass difference\footnote{The absolute neutrino masses and hierarchy of the neutrino mass splittings are not known at this time.} measured through oscillation experiments.  A reference for the performance of the energy resolution of the calorimeter under development for PTOLEMY is found here~\cite{portesi2015fabrication}.

The configurations of the planar geometry current coils and filter voltages are described below.

\subsection*{Setting of Magnetic Field Gradient}
\addcontentsline{toc}{subsection}{Setting of Magnetic Field Gradient}

\begin{figure*}[h!]
\begin{center}
      \includegraphics[width=0.45\textwidth]{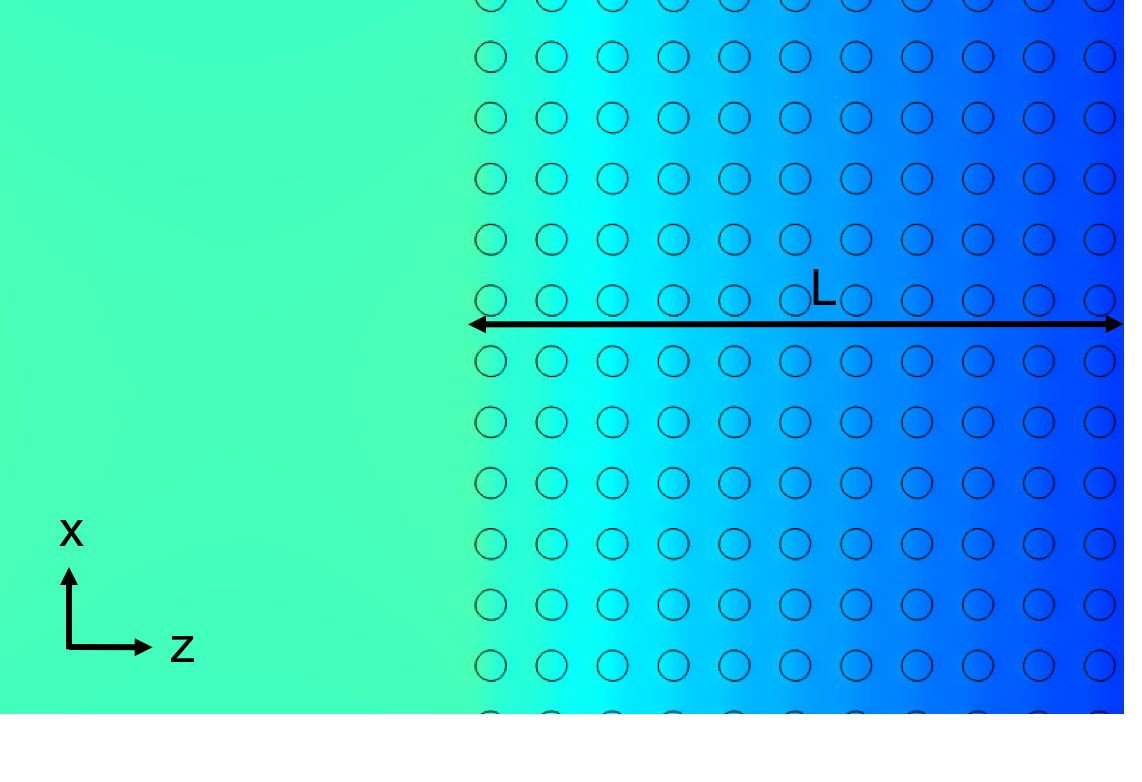} \qquad 
      \includegraphics[width=0.45\textwidth]{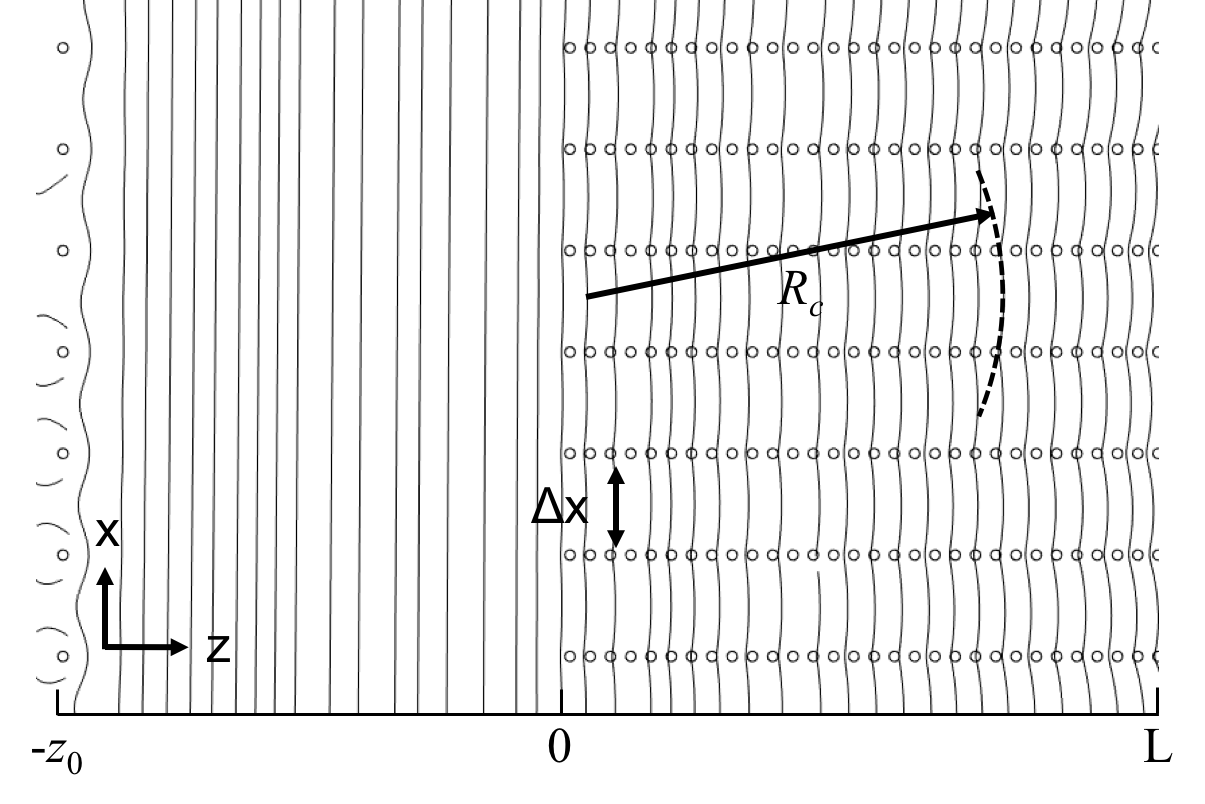}
   \caption{({\it left}) Tightly packed series of planes of current density over a distance $L$ in the $z$-direction, seen end-on and shown with circular cross-section wires, yielding an exponentially falling, straight $B_x$ field with $B_y=B_z \approx 0$, where shading is used to represent the total magnitude of the magnetic field from green (highest) to blue (lowest). ({\it right}) A finite $\Delta x$ gap between planes of current density introduces a radius of curvature, $R_c$, in the $B_x$ field, plotted as field lines, in the vacuum region between current coils.
   }
\label{fig:jint}
\end{center}
\vspace*{-15pt}
\end{figure*}

The configuration of a magnetic field gradient that optimally satisfies
the conditions for adiabatic invariance is such that
\begin{equation}
\left| B/\nabla{B} \right| \sim \epsilon \rho_c = \frac{\epsilon}{q} \sqrt{\frac{2 m \mu}{B}} 
\end{equation}
where $\epsilon$ is a dimensionless factor with $\epsilon \ll 1$.
As an example, we choose an exponentially falling $B_x$ magnetic field component
with characteristic length scale $\lambda$, given by
\begin{equation} \label{eq:Bexpfall}
B_x(z) = B_0 e^{-z/\lambda} \ .
\end{equation}
This field can be approximated by a series of surfaces of current density flowing along the $y$-direction.  Starting with an infinite $x$-$y$ plane of linear current density $J_+$ [A/m] flowing in the $y$-direction that passes through the coordinate $-z_0$, a constant magnetic field with magnitude $B_0$ is generated in the $x$-direction.  Starting at $z=0$ and extending to $z=L$, a current density $j_-(z)$ flowing along the  negative $y$-direction has an exponential fall off in $z$
with characteristic length scale $\lambda$.  The current density $j_-$ [A/m$^2$] integrated over $z$ is equal and opposite to $J_+$
\begin{equation}
J_+ + \int_0^L j_-(z) dz = 0
\end{equation}
where $L$ is the finite thickness in $z$ of the volume of non-zero current density $j_-(z)$, as shown in Figure~\ref{fig:jint}.
For these ideal conditions, the $B_x$ component is zero for $z>L$ and $z<-z_0$ with
a constant magnitude, $B_0$, in the interval $-z_0<z<0$ and an exponential drop off for
$0 \le z \le L$, as described by equation~(\ref{eq:Bexpfall}).

We introduce vacuum separations
$\Delta x$ between planes of current in the $x$-direction, such that $\Delta x < \lambda$,
as shown in Figure~\ref{fig:jint}.
In vacuum, we have $\bm{\nabla} \times \bm{B} = 0$.
Applying this to the following, we evaluate the transverse component of the magnetic field gradient to give
\begin{equation}
\bm{\nabla_\perp} B =  \left( \bm{B \cdot \nabla} \right) \left( \frac{\bm{B}}{B} \right) - \left( \bm{\nabla} \times \bm{B} \right) \times \left( \frac{\bm{B}}{B} \right) = - \frac{B}{R_c} \bm{\hat{n}}
\end{equation}
where $R_c$ is the radius of curvature and $\bm{\hat{n}}$ is the unit vector normal to the magnetic field line curvature.  For the example given above for the exponentially falling $B_x(z)$
field component, the radius of curvature in the $y$-$z$ plane is equal to the length scale $\lambda$ ($R_c = \lambda$).
The magnetic field components satisfying the $\bm{\nabla} \times \bm{B} = 0$ vacuum conditions are approximated in the central gap region between 
planes of current coils\footnote{The current coils are envisioned to be wound in a planar, also known as pancake, geometry.} by
\begin{align}
B_x & = B_0 \cos \left( \frac{x}{\lambda} \right) e^{-z/\lambda} \ , \label{eq:bxfield} \\
B_y & = 0 \ , \label{eq:byfield} \\
B_z & = - B_0 \sin \left( \frac{x}{\lambda} \right) e^{-z/\lambda} \ . \label{eq:bzfield}
\end{align}
In the mid-plane of the $\Delta x$ gap ($x=0$ and repeated periodically), the normal to the curvature of the $\bm{B}$ field points along the $z$-direction, $\bm{\hat{n}} = \bm{\hat{z}}$.

This configuration of $\bm{B}$ field has been shown in numerical simulations using the Fast Multipole Method (FMM)~\cite{greengard1987fast} (Leslie Greengard, Simons Foundation) to yield over four orders of magnitude reduction in the $B_x$ field with $\mathrm{\lambda=0.05\,m}$.



\subsection*{Setting of Filter Voltages}
\addcontentsline{toc}{subsection}{Setting of Filter Voltages}

The filter will act in a static configuration on an electron entering the
electrodes, two $y$-$z$ planes of segmented conductors, at a given reference
voltage $V_0$ and height $y_0$ between the electrodes.  If the voltage difference across the top and bottom electrodes (in the $y$-direction)
is a constant as a function of $z$, the decaying $B_x$ magnitude will induce
an exponential increase in the $\bm{E} \times \bm{B}$ drift and through gradient-$B$ drift cause
the electron to drift in the $y$-direction.

\begin{figure*}[h!]
\begin{center}
      \includegraphics[width=0.7\textwidth]{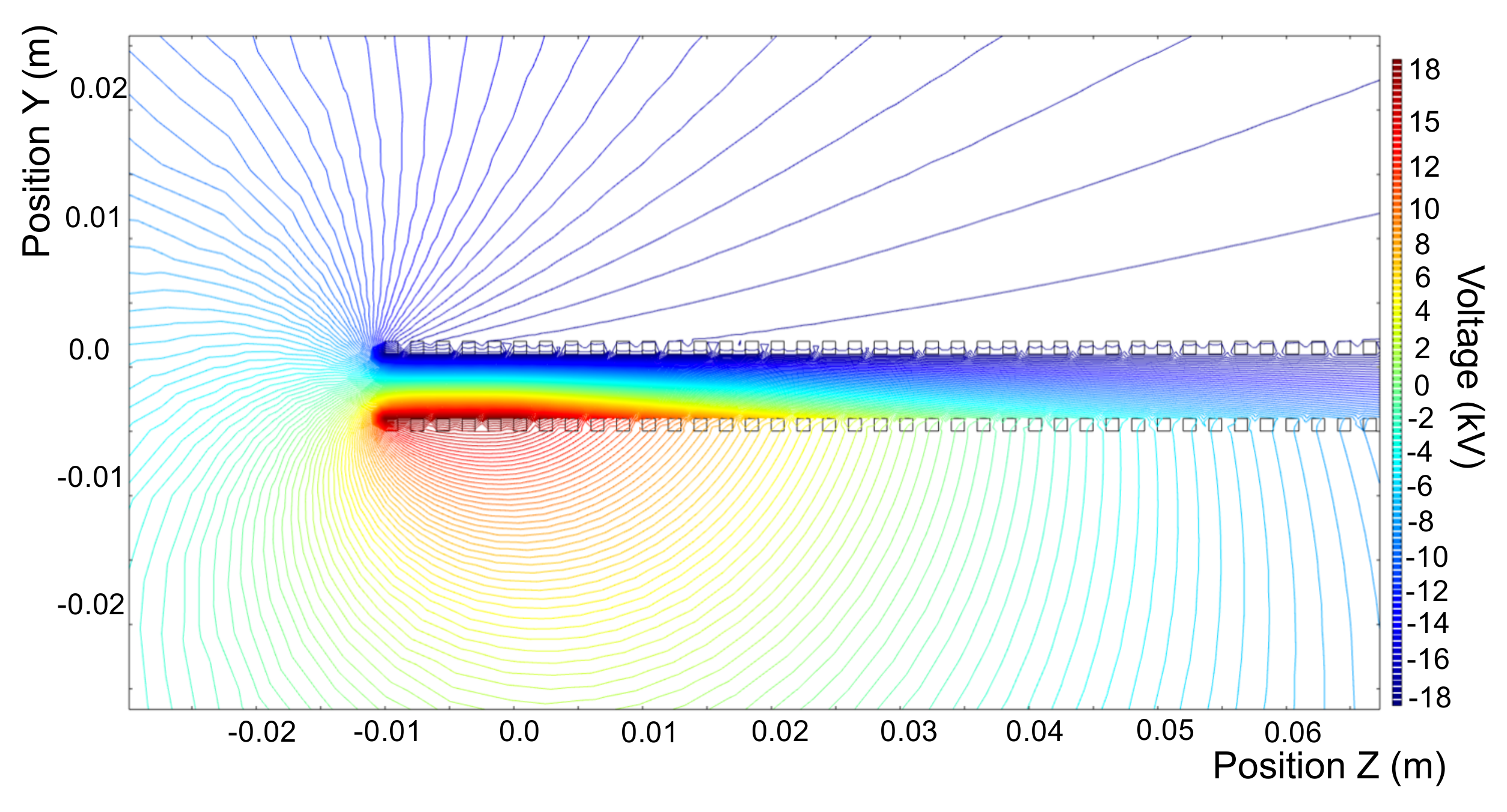}
   \caption{Example of filter voltage settings for a given input transverse kinetic energy and starting voltage level, computed with COMSOL software~\cite{multiphysics2015v}.  Here the voltage at the mid-distance between the plates in the vertical coordinate is set to $V_0=0$, the voltage of the top plate is held constant at $V_t = - \mu B_0$, and the voltages of the bottom segmented plates are set according to $V_b = \mu B_0 (2 e^{-z/\lambda} - 1)$ for $z \ge 0$.  The vertical separation of the plates in $y$ is small compared to the length scale $\lambda$ of the exponential ($y \ll \lambda$).
   }
\label{fig:voltages}
\end{center}
\vspace*{-15pt}
\end{figure*}

An alternative approach is to choose a voltage configuration that keeps the electron at a constant $y$-position as it travels down the length of the filter.  We propose a configuration where the voltage difference
across the plates in the $y$-direction is exponentially dropping with the same
characteristic length scale, $\lambda$, in the $z$-direction.  This is achieved
with electrodes segmented in the $z$-direction and yields a constant $\bm{E} \times \bm{B}$
drift.  To maintain the electron trajectory at a constant height of $y_0$, we
adjust the offset of the voltage difference across the plates as a function of $z$.
The combined set of configurations yields the following voltage, evaluated at the fixed
height $y=y_0$, as a function of $z$ for an initial voltage $V_0$ at $z=0$
\begin{equation} \label{eq:voltagecond}
V(y,z)|_{y=y_0} = V_0 - \mu B_0 \left( 1 - e^{-z/\lambda} \right) \ .
\end{equation}
An example voltage configuration is shown in Figure~\ref{fig:voltages} with $V_0=0$, the voltage of the top plate is held constant at $V_t = - \mu B_0$, and the voltages of the 
bottom segmented plates are set according to $V_b = \mu B_0 (2 e^{-z/\lambda} - 1)$ for $z \ge 0$.  The vertical separation of the plates in $y$ is small compared to the length scale $\lambda$ of the exponential ($y \ll \lambda$).

\begin{figure*}[h!]
\begin{center}
      \includegraphics[width=0.49\textwidth]{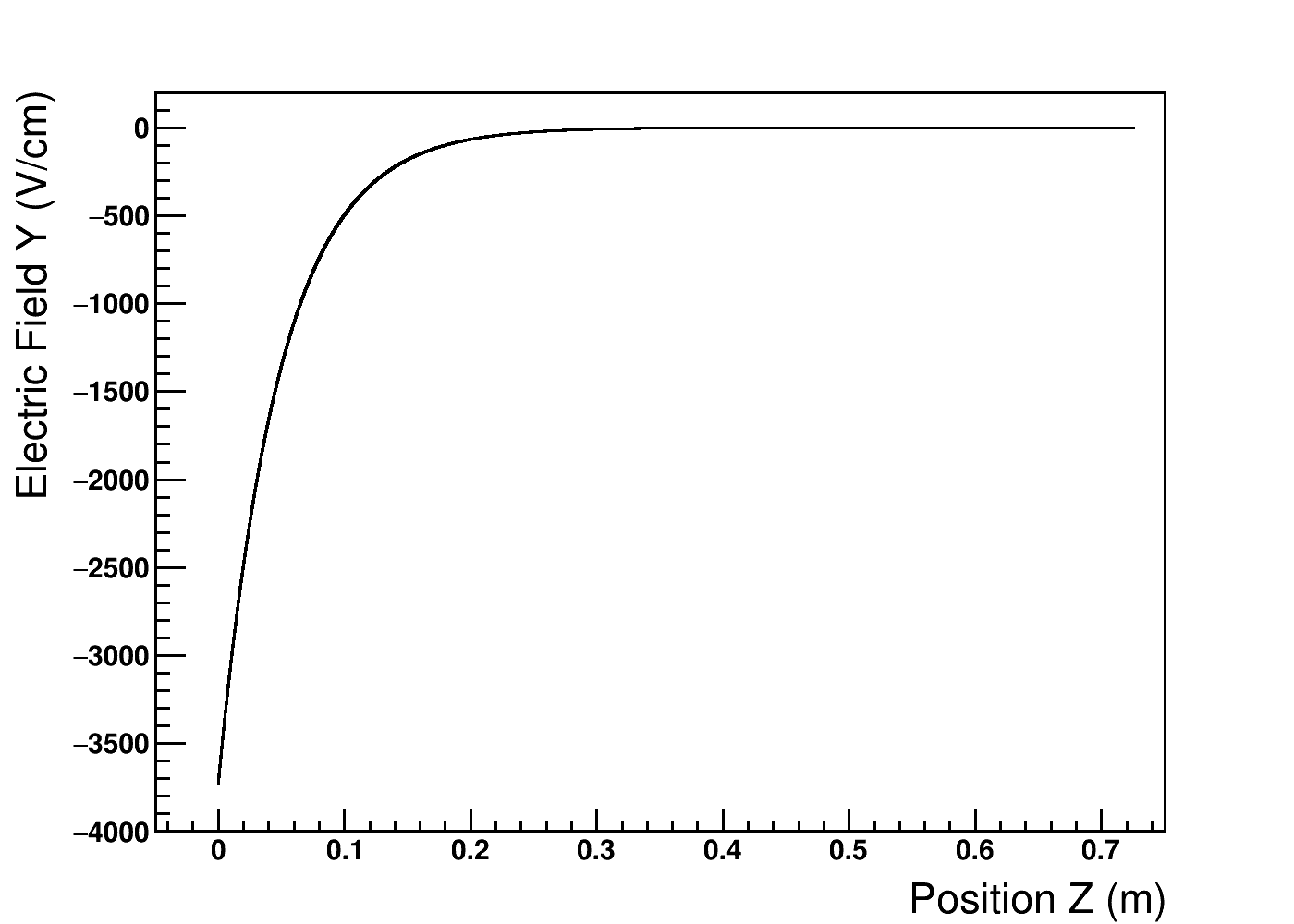}
      \includegraphics[width=0.49\textwidth]{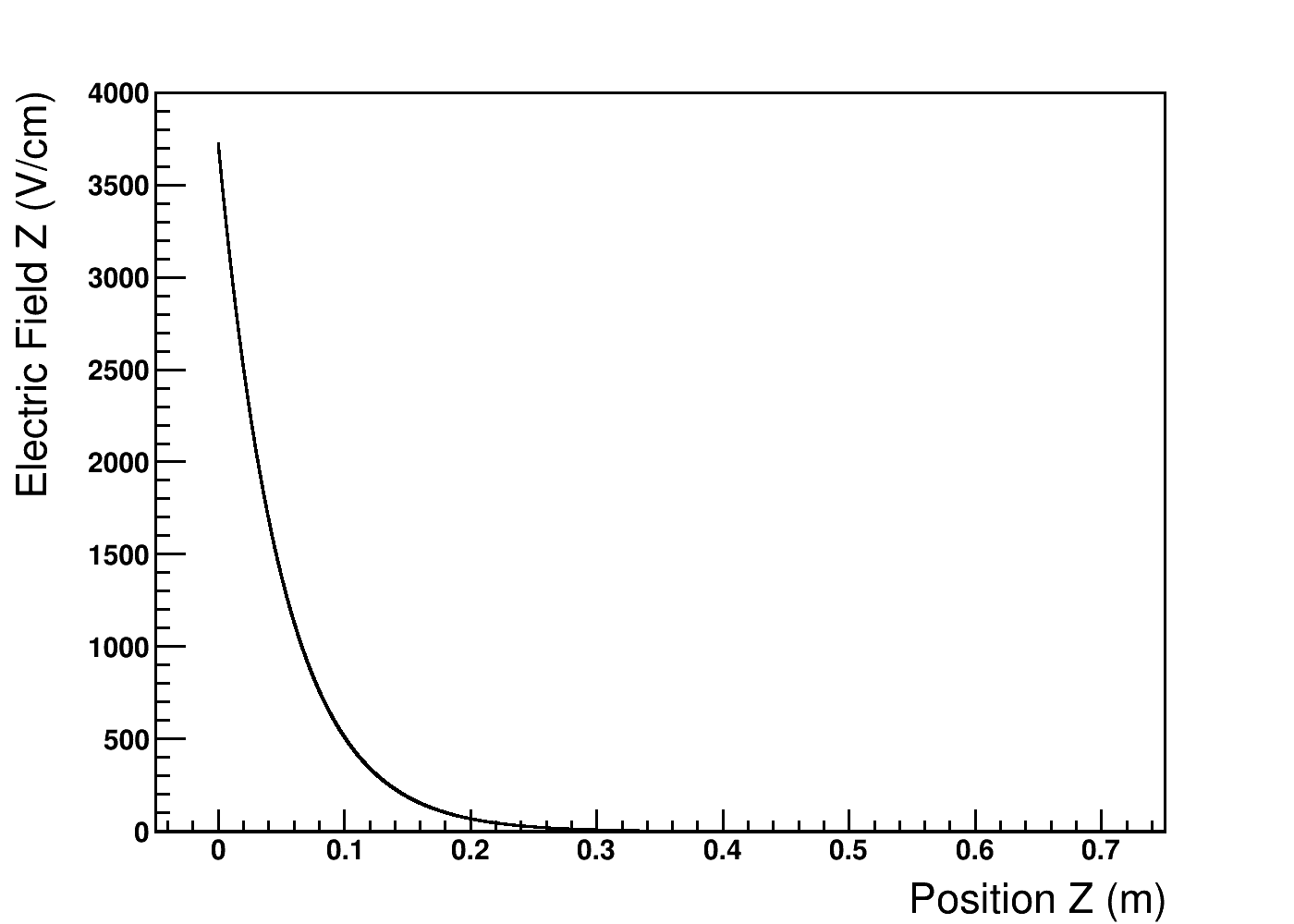}
      \includegraphics[width=0.49\textwidth]{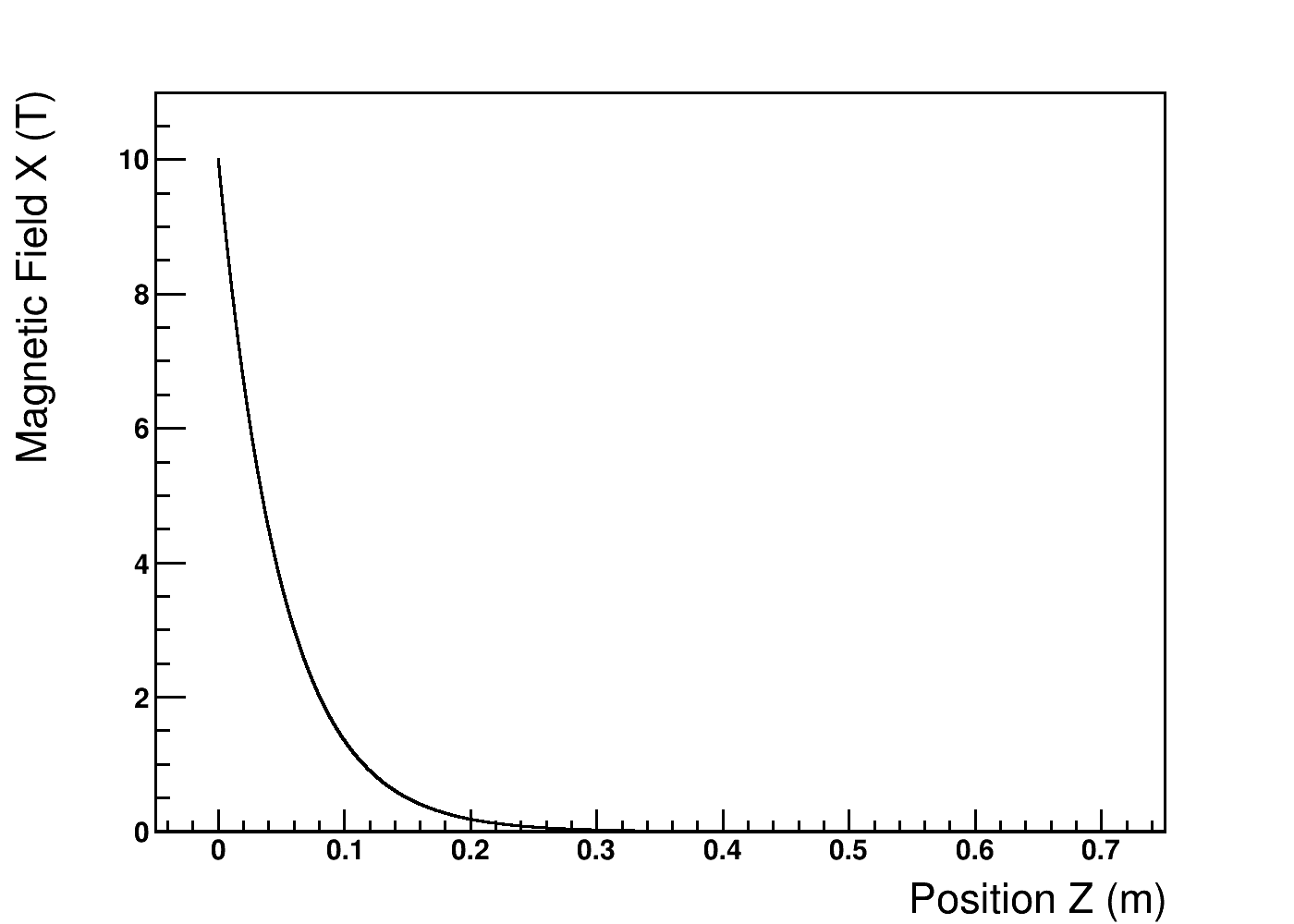}
      \includegraphics[width=0.49\textwidth]{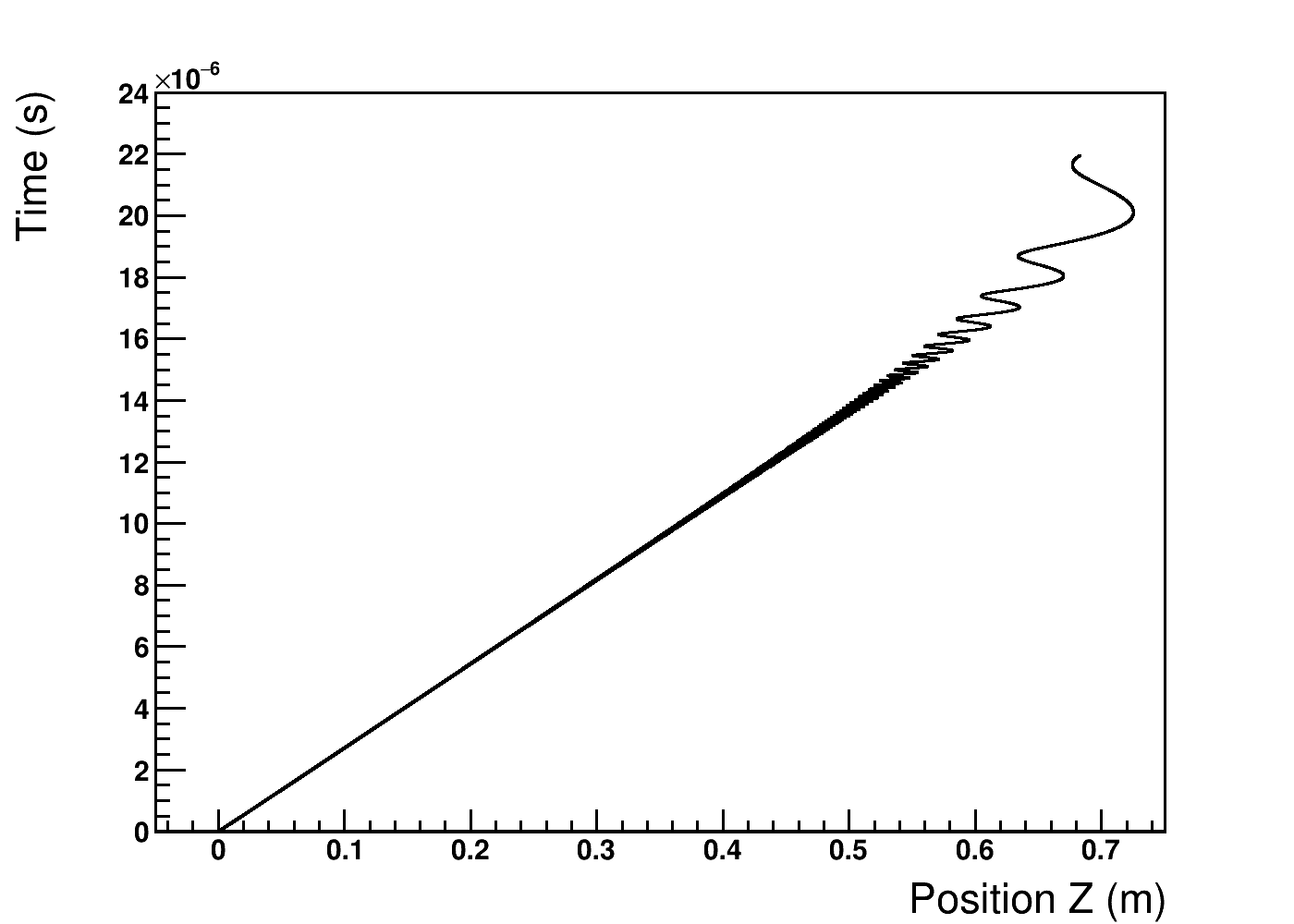}
   \caption{The $E_y$ ({\it top-left}) and $E_z$ ({\it top-right}) electric field components as a function of $z$ for an exponentially falling $B_x(z)$ ({\it bottom-left}) magnetic field.  The time versus $z$-position of the electron drift ({\it bottom-right}) shows a linear progression.}
\label{fig:fields}
\end{center}
\vspace*{-15pt}
\end{figure*}

The corresponding electric field components are given (for $z \ge 0$) by\footnote{The constant voltage plate in the example voltage configuration in Figure~\ref{fig:voltages} corresponds to $y=0$.}
\begin{align}
E_x & = 0 \ , \label{eq:exfield} \\
E_y & = E_0 \cos \left( \frac{y}{\lambda} \right) e^{-z/\lambda} \ , \label{eq:eyfield} \\
E_z & = - E_0 \sin \left( \frac{y}{\lambda} \right) e^{-z/\lambda} \ , \label{eq:ezfield}
\end{align}
where $E_0$ is the total magnitude of the $\bm{E}$ field at $z=0$.
The $\bm{E}$ field satisfies the $\bm{\nabla} \times \bm{E} = 0$ condition for static $\bm{B}$ fields (Faraday's law) and $\bm{\nabla \cdot E} = 0$ in the vacuum region between the filter plates.
The value of $E_0$ is determined by setting the magnitude of the vertical $\bm{E} \times \bm{B}$ drift equal and opposite to the magnitude of the gradient-$B$ drift for a given 
$y = y_0$, such that
\begin{align}
\frac{E_z}{B} & \approx - \frac{E_0}{B_0} \sin \left( \frac{y_0}{\lambda} \right) \\
              & = - \frac{\mu}{q B} \frac{\partial B_x}{\partial z} \\ 
              & = \frac{\mu}{q \lambda} \ ,
\end{align}
where the approximation is taken that $x \ll \lambda$.
The value of $E_0$ is therefore
\begin{equation}
E_0 = - \frac{\mu B_0}{q \lambda \sin \left( {y_0}/{\lambda} \right) } \ .
\end{equation}
This value of $E_0$ enforces the constancy of the cyclotron-averaged height of the electron trajectory throughout, since the $z$-dependence of $E_z$ matches that of $\partial B_x/\partial z$.
Similarly, from equation~(\ref{eq:eyfield}), $E_y$ at constant height $y=y_0$ has the same $z$-dependence as the $B_x$ field.  As a consequence, the magnitude of the 
$\bm{E} \times \bm{B}$ drift velocity is constant along the electron trajectory.
The $E_y$ and $E_z$ electric field components are shown in Figure~\ref{fig:fields} as a function of $z$ for an exponentially falling $B_x(z)$ magnetic field.  Also shown in Figure~\ref{fig:fields} is the time versus position of the electron drift.

The electron input to the filter, not described here, would follow lines of constant voltage into the filter from the left in a region of constant $\bm{B}$ field.

\subsection*{Motion Parallel to the Magnetic Field}
\addcontentsline{toc}{subsection}{Motion Parallel to the Magnetic Field}

For pitch angles less than 90 degrees, the electron will have a non-zero velocity component parallel to the magnetic field line.
The parallel velocity component of the electron kinetic energy is largely unaffected by the transverse drift motion through the filter.  Unlike the magnetic adiabatic collimation techniques, there is no rotation of the total velocity of the electron into the direction along the magnetic field.  We can therefore consider the motion parallel to the magnetic field to be independent of the reduction of the transverse velocity through gradient-$B$ and $\bm{E} \times \bm{B}$ drift.

One approach to maintain the electron motion within a finite range in the $x$-direction is to trap the electron within an electrostatic potential well.
The well potential can be constructed to leave the $E_y$ component of the electric field unchanged at the location of the mid-plane of the filter ($y=y_0$).  This can be done with an approximate harmonic potential that is formed between two linear, side-rail barriers that run along the length of the filter at positions $x=\pm D$ with respect to the $x=0$ nominal center of the parallel motion (and repeated periodically for each gap separated by $\Delta x$).

For an electric field falling off linearly from a line potential, we can approximate the center of the electric field as
\begin{equation} \label{eq:exharmonic}
E_x = \frac{k}{D+x} - \frac{k}{D-x} \approx 
\frac{k}{D} \left( (1 - x/D) - (1 + x/D) \right)
= - \frac{2k}{D^2}x
\end{equation}
where one can satisfy the $\bm{ \nabla \cdot E} = 0$ and 
$\bm{\nabla} \times \bm{E} = 0$ conditions with the additional components
\begin{align}
E_y & = \frac{2k}{D^2}(y - y_0) \ , \label{eq:eyharmonic} \\
E_z & = 0 \ , \label{eq:ezharmonic}
\end{align}
where $k$ sets the scale of the harmonic potential and the range of motion in the $x$-direction for a given magnitude of parallel velocity.
The electric field components from equations~(\ref{eq:exharmonic}), (\ref{eq:eyharmonic}), and (\ref{eq:ezharmonic}) add in superposition to the filter-voltage-defined electric field components from equations~(\ref{eq:exfield}), (\ref{eq:eyfield}), and (\ref{eq:ezfield}).  Consequently, the filter and side-rail voltages are defined to simultaneously solve the requirements of the top-bottom electrodes for balancing the gradient-$B$ drift against the $\bm{E} \times \bm{B}$ drift and to produce the harmonic trap with the side-rail barriers.  In particular, the voltage reference for the $x=0$ minimum of the harmonic trap will be set by the transverse drift requirements from the top-bottom electrode voltages.
For $y=y_0$, there is zero contribution from the harmonic trap to the drift $E$ field, as indicated from equations~(\ref{eq:eyfield}), (\ref{eq:ezfield}), (\ref{eq:eyharmonic}), and (\ref{eq:ezharmonic}).

\begin{figure*}[h!]
\begin{center}
      \includegraphics[width=0.85\textwidth]{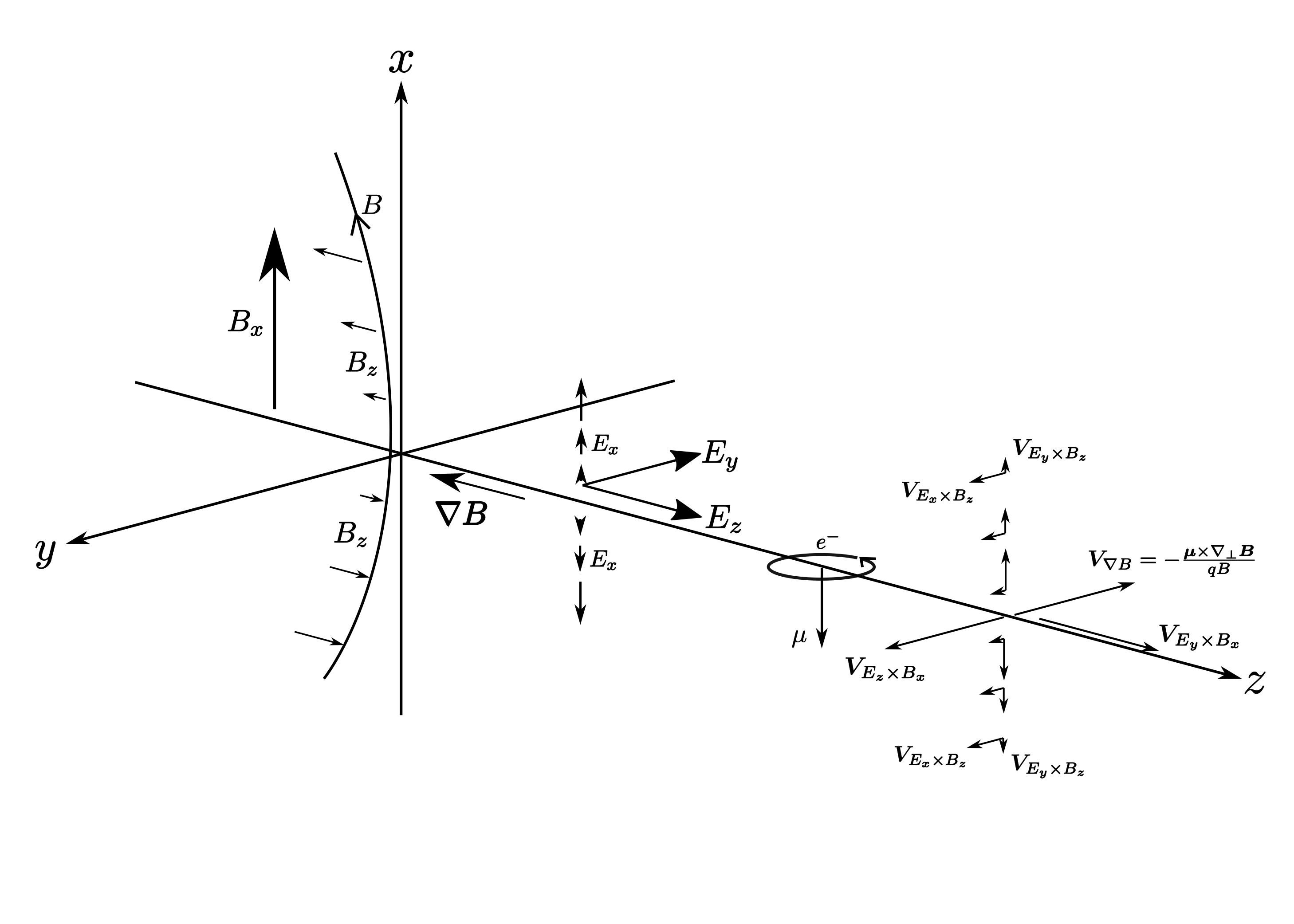}\\
      \includegraphics[width=0.7\textwidth]{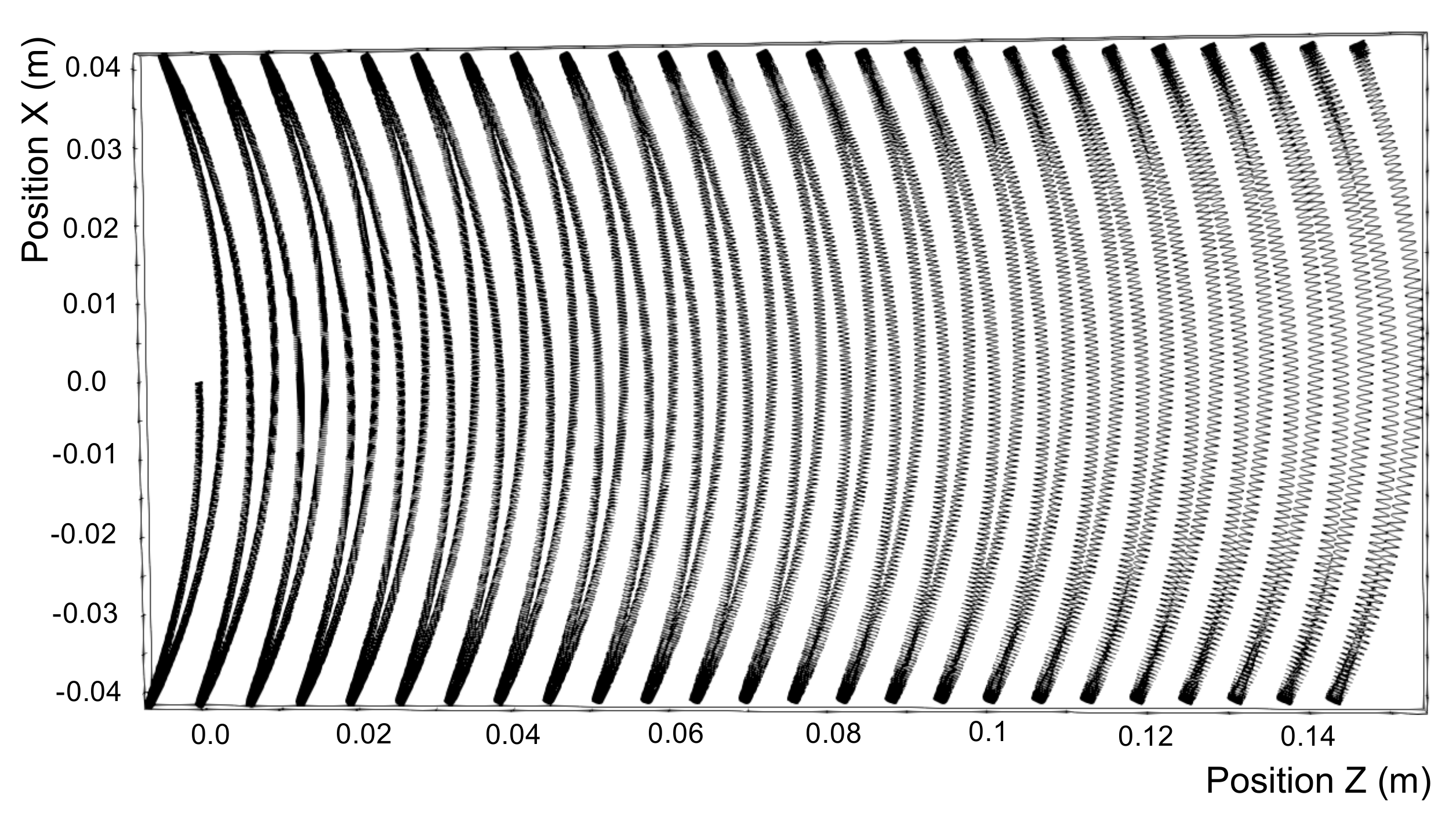}
   \caption{Configuration of magnetic and electric fields acting on the electron ({\it top}).  Parallel motion of an electron with an initial pitch angle of 85$^\circ$ in a harmonic trap ({\it bottom}).  For this example, we set $\lambda = 0.1$\,m and widen the distance along $x$ between side-rails to better visualize the motion.  The number of cyclotron revolutions per bounce decreases as the electron drifts into lower $B_x$ fields.
   \label{fig:parmotion}}
\end{center}
\vspace*{-15pt}
\end{figure*}

The configuration of magnetic and electric fields acting on the electron are shown in Figure~\ref{fig:parmotion}.
The parallel motion of an electron is shown in Figure~\ref{fig:parmotion} with the vacuum $\bm{B}$-field components defined in equations~(\ref{eq:bxfield}),
(\ref{eq:byfield}), and (\ref{eq:bzfield}).  The arc motion of the electron is a consequence of Amp\`ere's law applied to the exponentially decreasing $B_x$ field component.  The arc motion forces the electron to push into larger $\bm{B}$ fields at the turning points of the bound parallel motion.  Therefore, there will be mirror forces on the electron acting parallel to the $\bm{B}$ field in addition to the $\bm{E \cdot B}$ bottle forces from the electrostatic harmonic trap.

It should be noted that the electron motion in a harmonic trap induces a known frequency modulation, $f_m \sim \sqrt{2 k q/D^2 m}/2 \pi \sim 1$\,GHz for $kq \sim $\,18.6\,keV and $D \sim 1$\,cm, of the relativistic correction to the cyclotron frequency.  If sufficient RF measurement sensitivity is implemented for the estimation of $\mu$ in advance of entering the filter where the $\bm{B}$ field is uniform, this modulation may provide additional sensitivity on the estimation of the total kinetic energy of the electron, especially for the potential to phase-lock on the modulation frequency of the RF signal from cyclotron radiation emission.  In the filter region, where the $\bm{B}$ field is dropping exponentially, it is advantageous\footnote{The conditions of adiabatic invariance, given in equations~(\ref{eq:adinv1}) and (\ref{eq:adinv2}), rely on cyclotron-phase-averaged quantities.} to keep $f_m$ below the cyclotron frequency, $f_c = (1/2\pi) q B/m \gamma \sim $\,27\,GHz for $B \sim $\,1\,T and $\gamma \sim 1.03$.  The decreasing number of cyclotron orbits per bounce is shown in Figure~\ref{fig:parmotion}.  

One approach to reduce the frequency, $f_m$, of the bouncing motion for the same range of motion set by the side-rails at positions $x=\pm D$ is to reduce the magnitude of the parallel velocity of the electron and set the side-rail voltages in the filter to produce a shallower harmonic trap (smaller $|k|$).  The magnitude of the parallel velocity varies as the electron climbs the voltage potential of a harmonic trap.  In the region of constant magnetic field before entering the filter, a combination of $\bm{E} \times \bm{B}$ drift and parallel, bouncing motion can produce an electron trajectory that allows the electron to enter the filter with a lower parallel velocity and a higher (more negative) voltage reference $V_0$ at the center of the trap.  The details of this procedure are not included here.

\subsection*{Interface to the Calorimeter}
\addcontentsline{toc}{subsection}{Interface to the Calorimeter}

The parallel motion of the electron will be interfaced to the calorimeter measurement by setting a precision voltage on the calorimeter absorber 
to within a few Volts of the kinetic energy (divided by $q$) expected for electrons from the
tritium endpoint.  The estimate for the voltage of the calorimeter is determined from the estimate of $\mu$ from the RF signal and the corresponding starting values for the component of the kinetic energy from the perpendicular velocity
and solving for the complementary component of the kinetic energy from the
parallel velocity assuming the electron is originating from the tritium endpoint.

The calorimeter is located at the exit of the filter in the low magnetic field region (below mT fields).  The suppression of the kinetic energy perpendicular to the magnetic field is known as a function of length along the filter, and the suppression of the kinetic energy parallel to the magnetic field is known as a function of the $x$ displacement.  When combined, the total kinetic energy is known to within a few eV for a given location and reference voltage.
The calorimeter, located in the region of the side-rail, is set to a voltage to allow the electron to hit the surface. 
An array of calorimeters along the length of the exit of the filter will account for the range of estimated final kinetic energies and positions from the low field cyclotron motion.
 
The high resolution calorimeter measurement provides the final step to reach the target of 0.05\,eV total energy resolution in combination with a high precision voltage reference difference between the tritium target and calorimeter, with corrections for the RF cyclotron radiation emission losses.  Electrons which fall below the endpoint will not swing far enough in the $x$-direction to hit the calorimeter and will be dumped in an end electrode past the calorimeter region.  Electrons that are more energetic than the endpoint will have either escaped the harmonic trap and hit the side-rail or will hit the calorimeter and saturate the dynamic range of the energy measurement.  Electrons within a defined region of the endpoint, set by the dynamic range of the calorimeter, will have their energy spectrum measured over that region.  This technique provides differential spectral analysis of the endpoint and CNB signal region over the limited dynamic range of the calorimeter\footnote{The energy resolution of the calorimeter improves when the dynamic range is reduced.  The final choice of dynamic range will be chosen to match the filter performance.}.

\subsection*{Simulation of the Filter Performance}
\addcontentsline{toc}{subsection}{Simulation of the Filter Performance}

The electron trajectories in the filter are simulated with the Kassiopeia software package~\cite{furse2017kassiopeia}.

\begin{figure*}[h!]
\begin{center}
      \includegraphics[width=0.49\textwidth]{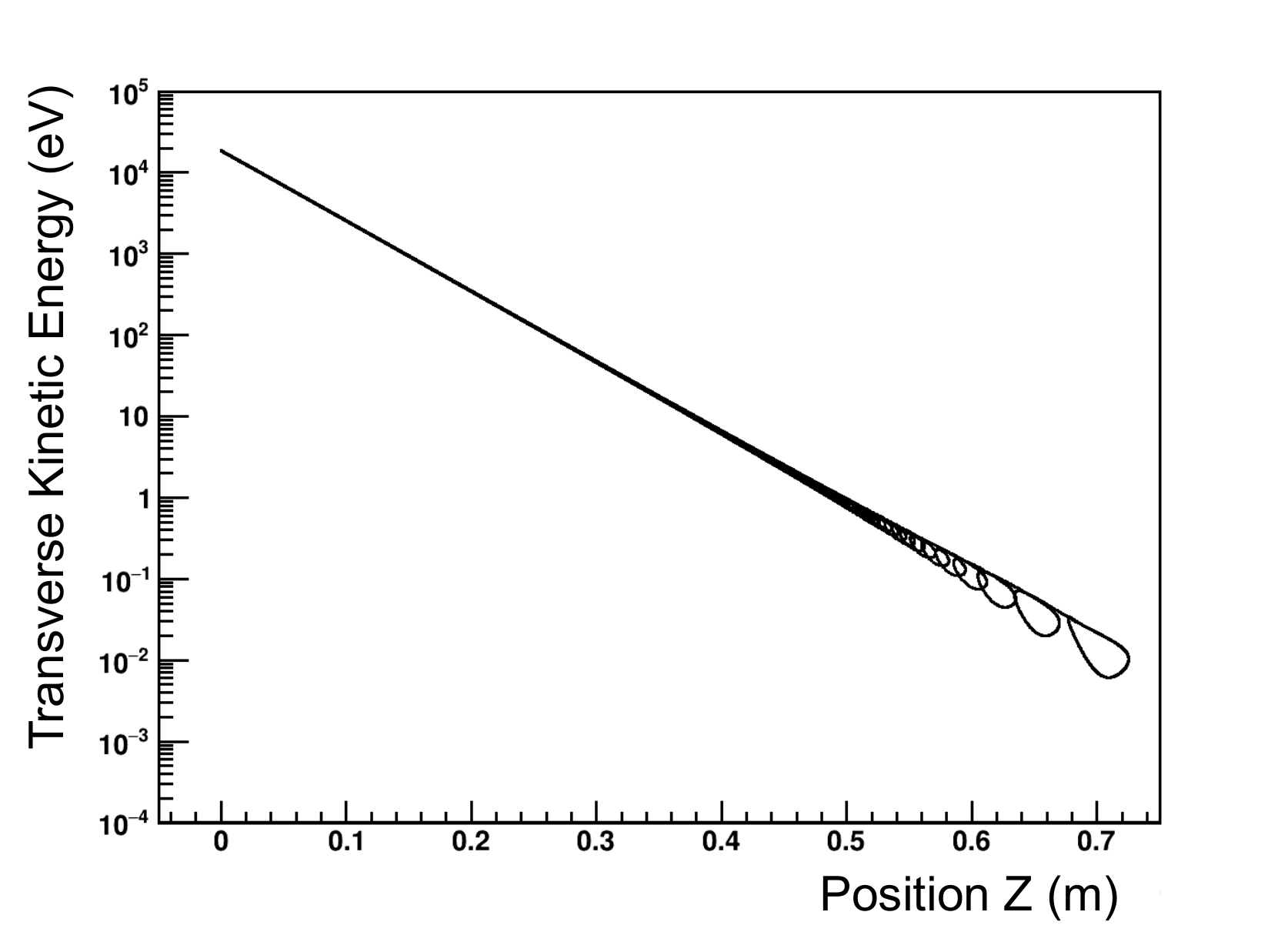}
      \includegraphics[width=0.49\textwidth]{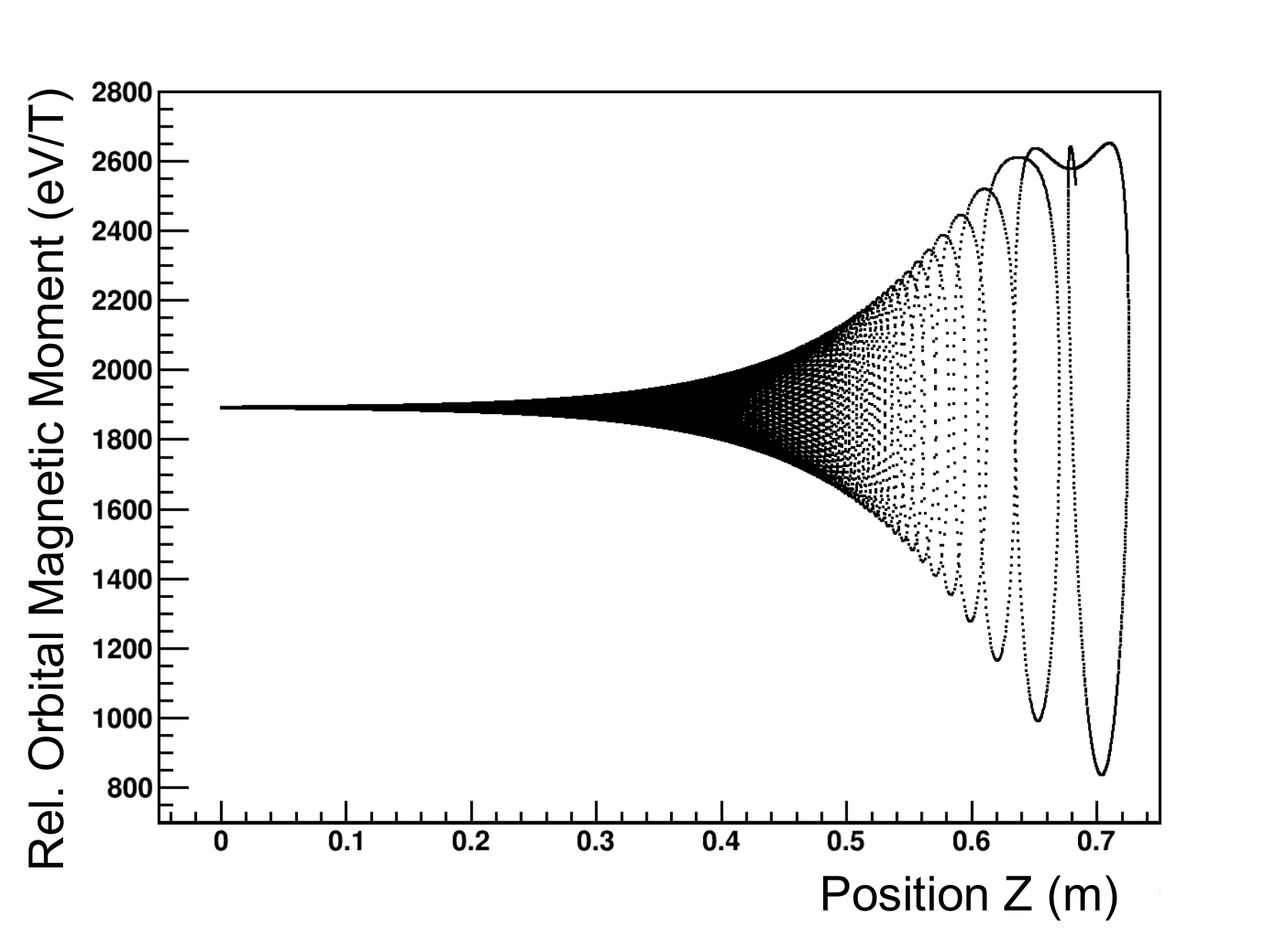}
   \caption{The transverse kinetic energy for an electron with a pitch angle of 90$^\circ$ (no parallel motion) and a starting value of 18.6\,keV at $z=0$ drops below 1\,eV after 0.5\,m in the filter with $\lambda =0.05$\,m ({\it left}).  The instantaneous value of the orbital magnetic moment [eV/T] displays oscillatory behavior from the electron drift for large $z$-distance along the filter length ({\it right}) while the cyclotron-averaged value of $\mu$ remains constant throughout the motion.}.
\label{fig:KEu}
\end{center}
\vspace*{-15pt}
\end{figure*}

Figure~\ref{fig:KEu} shows the kinematic quantities for a simulated electron trajectory with a starting transverse kinetic energy of 18.6\,keV at $z=0$ and a pitch angle of 90$^\circ$ (no parallel motion).  As the electron drifts from an initial high magnetic field region of 10\,T into the exponentially dropping low magnetic field for $z>0$, the cyclotron-averaged orbital magnetic moment remains constant, also shown in Figure~\ref{fig:KEu}.  Here we have set $\lambda =0.05$\,m in the exponential.  Therefore, the transverse kinetic energy drops with increasing $z$ and is below 1\,eV for a $z > 0.5$\,m.  This is where the interface to the calorimeter would be inserted.  Beyond $0.6$\,m, the instantaneous orbital magnetic moment has a strong oscillatory contribution from the electron drift velocity.

\begin{figure*}[h!]
\begin{center}
      \includegraphics[width=0.49\textwidth]{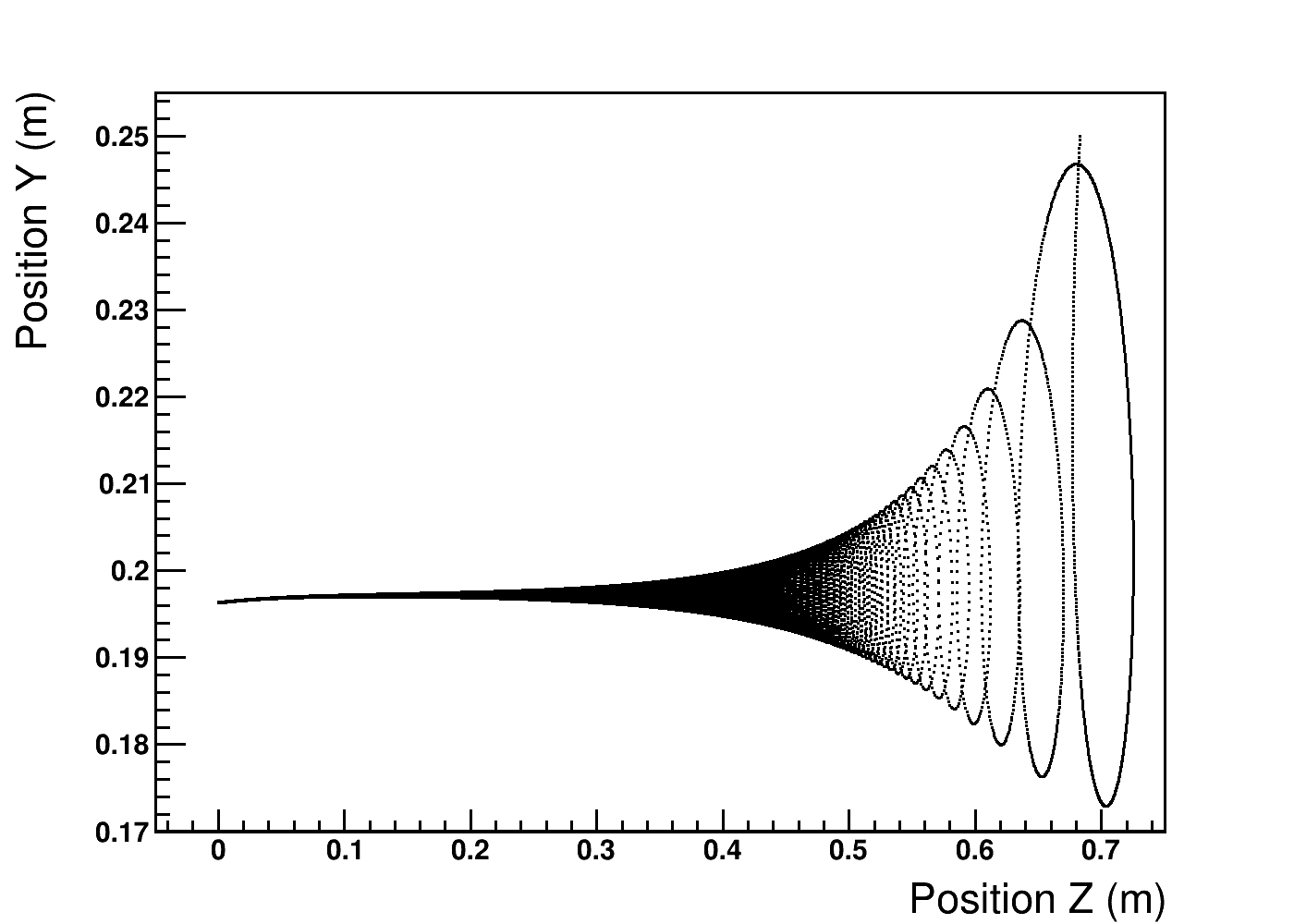}
      \includegraphics[width=0.49\textwidth]{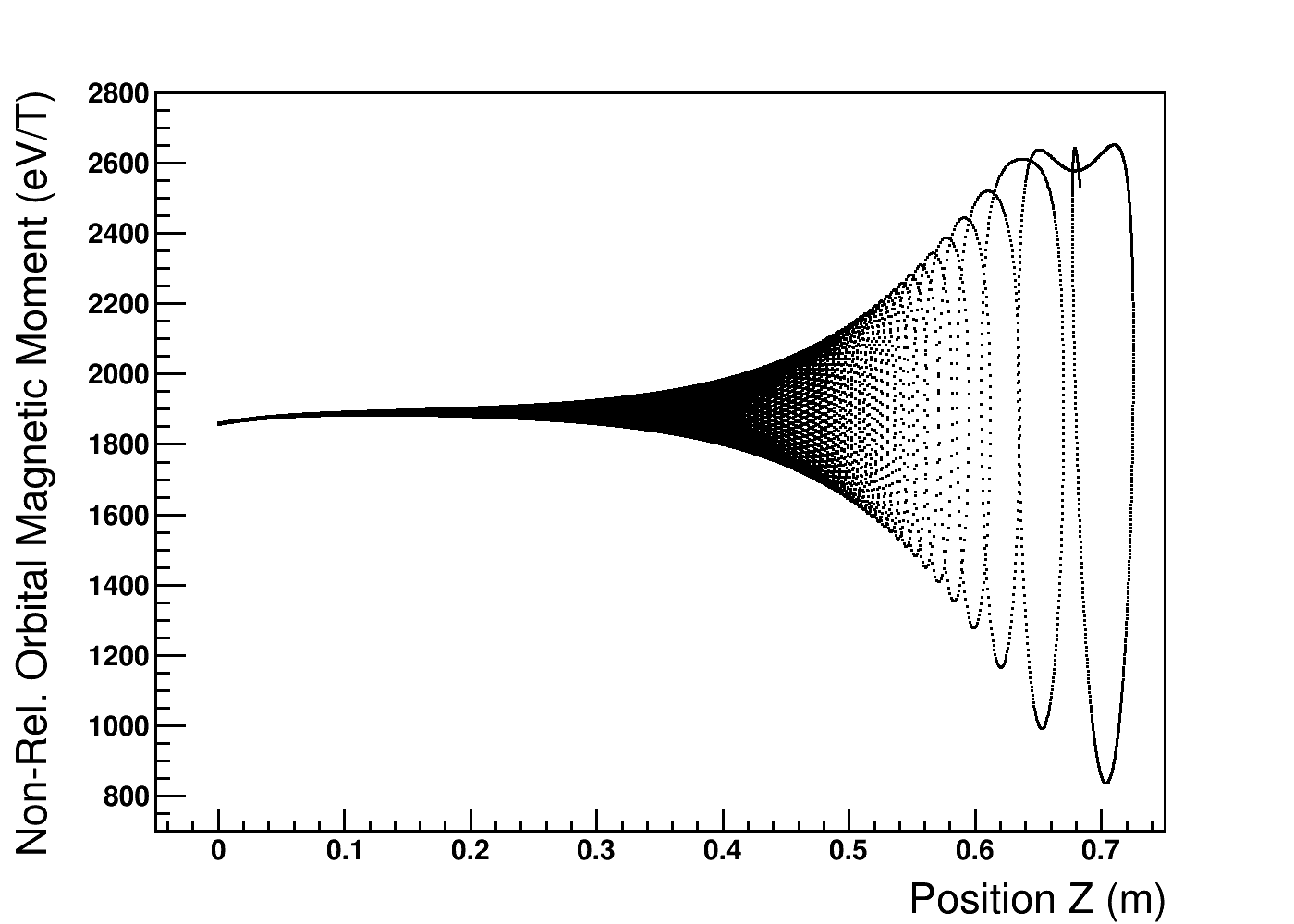}
   \caption{The $y$-position of the electron for the motion in Figure~\ref{fig:KEu} drifts slightly upwards in the semi-relativistic
   region at the start of the trajectory ({\it left}).  The same deviation is present in the non-relativistic expression for the orbital magnetic moment [eV/T] ({\it right}).}
\label{fig:yheight}
\end{center}
\vspace*{-15pt}
\end{figure*}

The $y$-position of the electron for the motion in Figure~\ref{fig:KEu} deviates from being perfectly constant over the filter trajectory due to semi-relativistic effects not included in equations~(\ref{eq:exfield}), (\ref{eq:eyfield}), and (\ref{eq:ezfield}) for the electric field values,
as shown in Figure~\ref{fig:yheight} near $z=0$.

If the true value of $\mu$ is less than the initial estimate from the RF signal, the electron will drift vertically downward and hit the bottom electrodes of the filter.  Similarly, values of $\mu$ larger than the estimate will drift vertically upwards and hit the top plate.  As the voltage range between the plates is decreasing exponentially with drift distance along the length of the filter, the sweeping away of electrons with neighboring $\mu$ values is exponential.  This process is the phase-space transformation of momentum into position space and is expected from Liouville's theorem for static Hamiltonian systems.  This also explains why the input area of the filter entrance can be equal to the area of the exit -- the bulk of the area expansion in the filtering process goes into electron trajectories that intercept the top and bottom plates.  The specific driving terms for the phase-space volume conservation are evaluated in the next section.

\section*{Conservation of Phase-Space Volume}
\addcontentsline{toc}{section}{Conservation of Phase-Space Volume}

The phase-space volume-preservation property of Hamiltonian systems~\cite{cary2009hamiltonian}, as follows from Liouville's theorem, 
can be expressed in the 
form\footnote{An equivalent formulation in terms of distribution functions is known as the Vlasov equation~\cite{roederer2014particle}.}:
\begin{equation} \label{eq:liouville}
0 = \frac{\partial B^*_\parallel}{\partial t} 
+ \bm{\nabla \cdot} (B^*_\parallel \bm{V})
+ \frac{\partial}{\partial p_\parallel} (B^*_\parallel \dot{p}_\parallel)
\end{equation}
where $\bm{V}$ is the GCS velocity and $p_\parallel$ is the momentum component of the electron parallel to the magnetic field direction, $\bm{\hat{b}} = \bm{B}/B$, and $\dot{p}_\parallel$ is the GCS force.  The effective magnetic field $\bm{B}^*$ for the calculation of the GCS phase space is given by 
\begin{equation}
\bm{B}^* \equiv \bm{B} + (c p_\parallel/q) \bm{\nabla} \times \bm{\hat{b}}
+ (m c/q) \bm{\nabla} \times \bm{V}_E
\end{equation}
with $c$ the speed of light, $\bm{V}_E$ the $\bm{E} \times \bm{B}$ drift velocity and, for what follows, 
$B^*_\parallel \equiv \bm{\hat{b} \cdot B}^*$~\cite{cary2009hamiltonian}.
The effective electric field for a static magnetic field is given by
\begin{equation}
\bm{E}^* \equiv - \bm{\nabla} \Phi^*
\end{equation}
with
\begin{equation}
q\Phi^* \equiv q \Phi + \mu B + (m/2)|\bm{V}_E|^2 \ .
\end{equation}
The effective potential directly incorporates $\mu$, which is proportional to the first adiabatic invariant~\cite{Alfven1940,cary2009hamiltonian}.

In equation~(\ref{eq:liouville}), the first term on the right is zero, as there is
no explicit time dependence, as seen by the electron, in the electromagnetic
fields of the filter.  Similarly, by applying Faraday's law, we can drop terms of the form $\bm{\nabla} \times \bm{E}$.
The static, non-relativistic terms contributing to the second and third terms are
given by
\begin{align} \label{eq:position}
\bm{\nabla \cdot} (B^*_\parallel \bm{V}) & = -c \bm{E}^* 
\bm{ \cdot \nabla} \times \bm{\hat{b}} \ , \ \mathrm{and} \\ \label{eq:momentum}
\frac{\partial}{\partial p_\parallel} (B^*_\parallel \dot{p_\parallel}) & = c \bm{E}^* \bm{\cdot \nabla} \times \bm{\hat{b}} \ ,
\end{align}
respectively.  These terms cancel identically in summation, as expected for phase-space volume conservation.  However, we can further explore the contributions to this term by expanding
\begin{equation}
\bm{\nabla} \times \bm{\hat{b}} = \left. \bm{\nabla} \times \bm{\hat{b}} \right|_\perp
+ \left. \bm{\nabla} \times \bm{\hat{b}} \right|_\parallel
\end{equation}
with
\begin{align}
\left. \bm{\nabla} \times \bm{\hat{b}} \right|_\perp & = \bm{\hat{b}} \times
\frac{\partial \bm{\hat{b}}}{\partial s} \ , \ \mathrm{and} \\
\left. \bm{\nabla} \times \bm{\hat{b}} \right|_\parallel & = \bm{\hat{b}} \,
\frac{\bm{\hat{b} \cdot} (\bm{\nabla} \times \bm{B})}{B} \ ,
\end{align}
where $s$ is the local coordinate along the GCS trajectory.  In vacuum, we have $\bm{\nabla} \times \bm{B} = 0$ and, therefore, only the term with $\partial \bm{\hat{b}}/{\partial s}$ contributes.
At the mid-plane of the $\Delta x$ gap, the normal to the curvature of the $\bm{B}$ field points along the $z$-direction, $\bm{\hat{n}} = \bm{\hat{z}}$, and the $\bm{B}$ field points along the $x$-direction.  By explicitly evaluating the phase-space movement term that appears in equations~(\ref{eq:position}) and (\ref{eq:momentum}) for the specific case of the transverse drift filter with the field configuration presented in Figure~\ref{fig:parmotion}, we find
\begin{equation} \label{eq:phasespace}
c \bm{E}^* \bm{ \cdot \hat{b}} \times
\frac{\partial \bm{\hat{b}}}{\partial s} = 
c (\bm{E} - \mu \bm{\nabla} B) \bm{\cdot \hat{x}} \times \left( - \frac{\bm{\hat{z}}}{R_c} \right)
= \frac{c}{R_c} \bm{E \cdot \hat{y}} 
- \frac{\mu c}{R_c} \bm{\nabla} B \bm{\cdot \hat{y}}
= \frac{c E_y}{R_c}
\end{equation}
where $\bm{\nabla} B = \bm{\nabla}_\perp B = - (B/R_c) \bm{\hat{z}}$
and therefore orthogonal to $\bm{\hat{y}}$.
Therefore, the non-zero term that actively shifts phase-space between position and momentum space to preserve the overall phase-space volume for the case of the transverse drift filter is precisely the product of the $E_y$ component of the electric field, against which the gradient-$B$ drift is doing work,
and the magnitude of $\nabla B$/$B$, given by $1/R_c$, times $c$.

In contrast, a magnetic collimation filter aligns the total velocity vector of the electron along the magnetic field, when moving under gyromotion from high magnetic field to low field, in the absence of electric fields.  The term responsible for phase-space volume conservation in a magnetic collimation filter is therefore the $\mu$ term that appears in equation~(\ref{eq:phasespace}).
The transverse drift filter is not a magnetic collimation filter and the $\mu$ term is null.

\section*{Conclusions}
\addcontentsline{toc}{section}{Conclusions}

We have presented the concept of the transverse drift filter for the study of the tritium endpoint spectrum and the measurement of the Cosmic Neutrino Background with PTOLEMY.  The filter differs from previous electromagnetic filters in that the motion of an electron emitted from a tritium target is driven by a combination of higher-order transverse drifts from a region of high magnetic field into low magnetic field.  We have studied the possibility of an exponentially falling magnetic field strength with a transverse drift filter that for a filter length of 0.7\,m reduces the transverse kinetic energy of an electron from a starting value of 18.6\,keV at the tritium endpoint down to 0.01\,eV, a scale that is comparable to neutrino masses.
The performance of the transverse drift filter is limited by the initial estimate of the transverse kinetic energy from a set of RF antennas that measure the cyclotron radiation emission from a single electron with a relativistically shifted cyclotron frequency, for tritium endpoint electrons, in advance of the electron entering the filter.  The final level of precision measurement is expected to occur in a calorimeter system interfaced to the exit of the filter at an energy range that matches the dynamic range of the calorimeter.
The question of where the expansion of position phase-space occurs for the corresponding reduction in momentum phase-space within the transverse filter is explained.  Electrons that enter the filter with a transverse kinetic energy that is outside the range for which the filter voltages have been configured are driven into the electrodes that produce the $\bm{E} \times \bm{B}$ drift.
An analysis of the phase-space volume invariance is presented and shown to obey Liouville's theorem for Hamiltonian systems.

\section*{Acknowledgments}
\addcontentsline{toc}{section}{Acknowledgments}

CGT is supported by the Simons Foundation (\#377485) and John Templeton Foundation (\#58851).

\bibliographystyle{h-physrev}

\bibliography{main} 

\begin{thebibliography}{10}

\bibitem{Weinberg:1962zza}
S.~Weinberg, {\em {Universal Neutrino Degeneracy}},
\newblock Phys. Rev. {\bf 128}, 1457 (1962).

\bibitem{Cocco:2007za}
A.~G. Cocco, G.~Mangano, and M.~Messina, {\em {Probing low energy neutrino
  backgrounds with neutrino capture on beta decaying nuclei}},
\newblock JCAP {\bf 06}, 015 (2007), hep-ph/0703075.

\bibitem{Betts:2013uya}
S.~Betts {\em et~al.},
\newblock {Development of a Relic Neutrino Detection Experiment at PTOLEMY:
  Princeton Tritium Observatory for Light, Early-Universe, Massive-Neutrino
  Yield},
\newblock in {\em {Proceedings, Community Summer Study 2013: Snowmass on the
  Mississippi (CSS2013): Minneapolis, MN, USA}}, 2013, arXiv:1307.4738.

\bibitem{Baracchini2018}
E.~Baracchini {\em et~al.},
\newblock {PTOLEMY: A Proposal for Thermal Relic Detection of Massive Neutrinos
  and Directional Detection of MeV Dark Matter},
\newblock 2018, arXiv:1808.01892.

\bibitem{hamilton1953upper}
D.~R. Hamilton, W.~P. Alford, and L.~Gross, {\em Upper limits on the neutrino
  mass from the tritium beta spectrum},
\newblock Physical Review {\bf 92}, 1521 (1953).

\bibitem{bergkvist1972high}
K.-E. Bergkvist, {\em {A high-luminosity, high-resolution study of the
  end-point behaviour of the tritium $\beta$-spectrum (I). Basic experimental
  procedure and analysis with regard to neutrino mass and neutrino
  degeneracy}},
\newblock Nuclear Physics B {\bf 39}, 317 (1972).

\bibitem{lubimov1980estimate}
V.~Lubimov, E.~Novikov, V.~Nozik, E.~Tretyakov, and V.~Kosik, {\em {An estimate
  of the $\nu_{\rm e}$ mass from the $\beta$-spectrum of tritium in the valine
  molecule}},
\newblock Physics Letters B {\bf 94}, 266 (1980).

\bibitem{Beamson1980Collimating}
G.~Beamson, H.~Q. Porter, and D.~W. Turner, {\em {The collimating and
  magnifying properties of a superconducting field photoelectron
  spectrometer}},
\newblock Journal of Physics E: Scientific Instruments {\bf 13}, 64 (1980).

\bibitem{Lobashev1985Method}
V.~Lobashev and P.~Spivak, {\em {A method for measuring the electron
  antineutrino rest mass}},
\newblock Nuclear Instruments and Methods in Physics Research Section A:
  Accelerators, Spectrometers, Detectors and Associated Equipment {\bf 240},
  305 (1985).

\bibitem{Picard1992Solenoid}
A.~Picard {\em et~al.}, {\em {A solenoid retarding spectrometer with high
  resolution and transmission for keV electrons}},
\newblock Nuclear Instruments and Methods in Physics Research Section B: Beam
  Interactions with Materials and Atoms {\bf 63}, 345 (1992).

\bibitem{kraus2005final}
C.~Kraus {\em et~al.}, {\em Final results from phase II of the Mainz neutrino
  mass searchin tritium $\beta$ decay},
\newblock The European Physical Journal C-Particles and Fields {\bf 40}, 447
  (2005).

\bibitem{aseev2011upper}
V.~Aseev {\em et~al.}, {\em Upper limit on the electron antineutrino mass from
  the Troitsk experiment},
\newblock Physical Review D {\bf 84}, 112003 (2011).

\bibitem{wolf2010katrin}
J.~Wolf {\em et~al.}, {\em The KATRIN neutrino mass experiment},
\newblock Nuclear Instruments and Methods in Physics Research Section A:
  Accelerators, Spectrometers, Detectors and Associated Equipment {\bf 623},
  442 (2010).

\bibitem{roederer2014particle}
J.~Roederer and H.~Zhang,
\newblock Particle fluxes, distribution functions and violation of invariants,
\newblock in {\em Dynamics of Magnetically Trapped Particles}, pp. 89--122,
  Springer, 2014.

\bibitem{Alfven1940}
H.~Alfv\'en, {\em {On the motion of a charged particle in a magnetic field}},
\newblock Ark. Mat. Astron. Fys. {\bf 25B}, 29 (1940).

\bibitem{cary2009hamiltonian}
J.~R. Cary and A.~J. Brizard, {\em Hamiltonian theory of guiding-center
  motion},
\newblock Reviews of Modern Physics {\bf 81}, 693 (2009).

\bibitem{project8}
D.~M. Asner {\em et~al.}, {\em Single-Electron Detection and Spectroscopy via
  Relativistic Cyclotron Radiation},
\newblock Physical Review Letters {\bf 114}, 162501 (2015).

\bibitem{multiphysics2015v}
C.~Multiphysics, {\em v. 5.2},
\newblock COMSOL AB, Stockholm, Sweden  (2015).

\bibitem{portesi2015fabrication}
C.~Portesi, E.~Taralli, L.~Lolli, M.~Rajteri, and E.~Monticone, {\em
  Fabrication and characterization of fast TESs with small area for single
  photon counting},
\newblock IEEE Transactions on Applied Superconductivity {\bf 25}, 1 (2015).

\bibitem{greengard1987fast}
L.~Greengard and V.~Rokhlin, {\em A fast algorithm for particle simulations},
\newblock Journal of Computational Physics {\bf 73}, 325 (1987).

\bibitem{furse2017kassiopeia}
D.~Furse {\em et~al.}, {\em {Kassiopeia: a modern, extensible C++ particle
  tracking package}},
\newblock New Journal of Physics {\bf 19}, 053012 (2017).

\end{thebibliography}
\addcontentsline{toc}{section}{References}


\end{document}